\documentclass[aps,prb,twocolumn,groupedaddress,amsmath,floatfix]{revtex4-2}

\usepackage{graphicx}
\usepackage{xcolor}
\usepackage{amssymb}

\newcommand*\diff{\mathop{}\!\mathrm{d}} 

\begin{document}

\title{Scalar susceptibility of a diluted classical XY model}
\author{Reece Beattie-Hauser and Thomas Vojta}
\affiliation{Department of Physics, Missouri University of Science \& Technology,\\ Rolla, MO, 65409, USA}

\date{\today}

\begin{abstract}
We analyze the amplitude fluctuations in a diluted 3D classical XY model near the magnetic phase transition, motivated by the unusual localization properties of the
amplitude (Higgs) mode recently found at the disordered superfluid-Mott glass quantum phase transition.
We calculate the amplitude correlation function and the corresponding scalar susceptibility by means of Monte Carlo simulations.
In contrast to the quantum case, in which the scalar susceptibility was found to violate naive scaling, we find that the scalar susceptibility of the classical system
fulfills naive scaling (employing the clean critical exponents, as expected from the Harris criterion) as the temperature is varied across the phase transition for
several dilutions. We discuss possible reasons for this discrepancy as well as the generality of our findings.
\end{abstract}

\maketitle

\section{Introduction}
In systems featuring a spontaneously broken continuous symmetry, the collective fluctuations about the ordered state can be classified into oscillations of the order
parameter direction and oscillations of the order parameter amplitude (for reviews, see Refs.\ \cite{PekkerVarma15,Burgess00}). Examples of continuous symmetry breaking
in condensed matter occur in planar or Heisenberg magnets, superfluids, superconductors, or optical lattice bosons. The direction oscillations, called the
Goldstone modes, are gapless (massless) as a consequence of Goldstone's theorem \cite{Nambu60,Goldstone61,GoldstoneSalamWeinberg62}, at least in the case of short-range
interactions. The amplitude oscillations, in contrast, are gapped (massive) and can be understood as the condensed matter analog of the Higgs boson \cite{Higgs64} in
particle physics. They are therefore often called Higgs modes.

In recent years, the behavior and observability of the amplitude (Higgs) mode in condensed matter has attracted considerable attention. In systems with Lorentz-invariant
low-energy dynamics, the amplitude and direction (phase) degrees of freedom decouple. Thus, a well-defined amplitude mode can exist provided it cannot decay rapidly
into lower-energy excitations \cite{PekkerVarma15}. Analytical and numerical studies of a relativistic $O(N)$ field theory
\cite{AffleckWellman92,PodolskyAuerbachArovas11,GazitPodolskyAuerbach13} have demonstrated that the amplitude mode is characterized by a pronounced spectral peak in the
scalar susceptibility (the susceptibility associated with the amplitude-amplitude correlations). This peak survives, in both two and three space dimensions, all the way
to the quantum critical point at which the symmetry-broken phase is destroyed. The peak energy (i.e., the Higgs mass) $\omega_H$ softens with decreasing distance $t$
from criticality, governed by the power-law $\omega_H \sim |t|^{\nu z}$ where $\nu z$ is the correlation time critical exponent.

These results apply to the clean, translationally invariant case. In the presence of quenched randomness, the character of the amplitude mode changes qualitatively.
Puschmann et al.\ investigated a site-diluted particle-hole symmetric quantum rotor model (in the same universality class as an $O(N)$ field theory with random-mass
disorder) by means of Monte Carlo simulation and an inhomogeneous mean-field theory \cite{PuschCrewseHoyosVojta20,CrewseVojta21,PuschGetelinaHoyosVojta21}. They
found that the Higgs spectral peak in the scalar susceptibility is absent for any nonzero dilution. Instead, the scalar response is characterized by a broad ``hump''
whose maximum is at some microscopic energy. Moreover, the scalar response is non-critical, i.e., it does not change appreciably as the system is tuned through the
quantum phase transition. This behavior, which violates naive scaling, suggests that the amplitude mode is spatially localized. However, the reasons for this
localization and the conditions under which it appears are not fully understood. Is the amplitude mode localization simply a manifestation of Anderson localization due
to the randomness in the (bare) Hamiltonian? Or is it related to renormalization phenomena, i.e., to the fact that the clean and disordered $O(N)$ field theories belong
to different universality classes whose critical exponents differ significantly from each other?

In the present paper, we therefore investigate the scalar susceptibility of a site-diluted three-dimensional \emph{classical} XY model to gain further understanding of
the amplitude fluctuations in a disordered system. For zero dilution, this XY model is described by the same field theory as the $(2+1)$-dimensional (particle-hole
symmetric) quantum rotor model studied in Refs.\ \cite{PuschCrewseHoyosVojta20,CrewseVojta21,PuschGetelinaHoyosVojta21} where the third dimension represents
imaginary time. Consequently, the classical phase transition in the clean XY model and the quantum phase transition in the clean rotor model are in the same universality
class. In contrast, the effects of nonzero dilution on the two systems differ from each other: According to the Harris criterion \cite{Harris74}, the disorder is
expected to be an irrelevant perturbation in the classical XY model whereas it is relevant in the quantum rotor model. In accordance, recent Monte Carlo simulations have
shown that the quantum phase transition in the diluted rotor model falls into a novel universality class \cite{VojtaCrewsePuschmannArovasKiselev16,ProkofevSvistunov04}.
Studying the scalar susceptibility of the diluted classical XY model and comparing it to that of the quantum rotor model will therefore help us to disentangle possible
causes for the amplitude mode localization discussed above.

Our paper is organized as follows. Section \ref{sec:model} introduces the model and its observables. The scalar susceptibility and the tools for its analysis are described in
Sec.~\ref{sec:scalar}. The technical specifications for our computer simulations are outlined in Sec.~\ref{sec:sim}. Section \ref{sec:results} contains our results for
the model's thermodynamics and the behavior of the amplitude fluctuations. The paper concludes in Sec.~\ref{sec:conc}.

\section{Three-dimensional site-diluted XY model}\label{sec:model}

We will be focusing on a classical site-diluted XY model on a cubic lattice, according to the Hamiltonian
\begin{equation}
H=-J\sum_{\langle ij \rangle} \epsilon_i \epsilon_j \mathbf{S}_i \cdot \mathbf{S}_j
\end{equation}
which is a sum over pairs of nearest-neighbor sites. Each $\mathbf{S}$ is a two-component unit vector spin, and each $\varepsilon$ is a quenched random variable, with:
\begin{equation}
  \epsilon = \left\{ \begin{array}{ll}
      0 & \; \; \; \mbox{with probability $p$} \\
      1 & \; \; \; \mbox{with probability $1-p$}
    \end{array}
    \right.
\end{equation}
where $p$ is the dilution concentration of a given system. In addition, we let $J$ be equal to unity.

The clean, undiluted XY model is known to undergo a magnetic phase transition at some critical temperature $T_c$ separating an ordered phase ($T<T_c$) and a
disordered phase ($T>T_c$). The system's behavior near $T_c$ has been studied in detail in the literature. Precise numerical values for the critical exponents were
computed, e.g., in Ref.~\cite{CampostriniHasenbuschPelissettoVicari06}. These include the correlation length critical exponent, which was found to be $\nu=0.6717$.

With increasing dilution $p$, the critical temperature is expected to decrease, reaching zero at the lattice percolation threshold $p_c$ (which takes the value of
$p_c=0.6884$ in the cubic lattice~\cite{Ballesterosetal99}). The critical behavior in the presence of dilution can be predicted through the Harris criterion, which
states that the diluted system's critical behavior will not change from the clean critical behavior if the clean correlation length critical exponent $\nu$ fulfills the
inequality
\begin{equation}\label{eq:harris}
  \nu d>2
\end{equation}
where $d$ is the dimensionality of the system. Since this is a three-dimensional model, the Harris criterion is fulfilled so we expect the diluted system to feature the
same critical exponents as in the clean case.

However, it should be noted that the prediction of the Harris criterion applies to the asymptotic critical behavior, i.e., the limit of infinite system size. We
emphasize that the inequality (\ref{eq:harris}) is fulfilled only \emph{barely}. This implies that the disorder strength scales to zero very slowly with increasing
system size, and a strongly diluted system will only reach the clean critical behavior at very large system sizes. These powerful finite size effects will be
demonstrated in more detail in Sec.~\ref{sec:results}.

To analyze the thermodynamics of our system near criticality, we make use of the Binder cumulant
\begin{equation}
  g=\left[1-\frac{\langle |\mathbf{m}|^4 \rangle}{3 \langle |\mathbf{m}|^2 \rangle^2}\right]_{\mathrm{dis}}
\end{equation}
where the angle brackets denote the canonical (Monte Carlo) average, the square brackets denote the disorder average, and $\mathbf{m}$ is the order parameter
(magnetization), defined as
\begin{equation}
  \mathbf{m}=\frac{1}{N} \sum_{i} \epsilon_i \mathbf{S}_{i}
\end{equation}
with $N=L^3$ being the number of lattice sites. The Binder cumulant is a dimensionless quantity; it thus has the scaling form $g(t,L)=Y(tL^{1/\nu})$, which implies that systems with different
linear system sizes $L$ will have intersecting Binder cumulant curves when the reduced distance from criticality $t=(T-T_c)/T_c$ is zero. Using this fact, we can calculate the
critical temperature $T_{c}$ by simply finding the intersection between different $g$ vs. $T$ curves. Additionally, the scaling form also provides a method for finding
the correlation length critical exponent $\nu$. The $g$ vs. $t$ curves for different $L$ can be collapsed into a single master curve by scaling $t$ by a constant factor
$X(L)$. These scaling factors can then be used to calculate $\nu$ using the power-law relationship
\begin{equation}\label{eq:powerlaw_form}
  X(L)\sim L^{1/\nu}.
\end{equation}

\section{Scalar susceptibility}\label{sec:scalar}

To examine the behavior of the Higgs mode, we compute the scalar correlation function, i.e., the correlation function of the order parameter amplitude,
\begin{equation}
  \begin{split}
    \chi_{\rho\rho}(\mathbf{x}) &= \left[\frac{1}{N} \sum_{\mathbf{x}'} \langle\rho(\mathbf{x}+\mathbf{x}')\rho(\mathbf{x}')\rangle\right. \\
                                &\hphantom{=[}\left.- \frac{1}{N} \sum_{\mathbf{x}'} \langle\rho(\mathbf{x}+\mathbf{x}')\rangle\langle\rho(\mathbf{x}')\rangle\right]_{\mathrm{dis}} \\
                                &= \left[\langle \rho(\mathbf{x}) \rho(\mathbf{0}) \rangle - \langle \rho(\mathbf{x}) \rangle\langle \rho(\mathbf{0})
                                \rangle\right]_{\mathrm{dis}}
  \end{split}
\end{equation}
as well as its Fourier transform, the scalar susceptibility,
\begin{equation}\label{eq:chi_fourier}
  \tilde{\chi}_{\rho\rho}(\mathbf{q})=\int \diff\mathbf{x}\, e^{-i\mathbf{q\cdot x}}\chi_{\rho\rho}(\mathbf{x})
\end{equation}
where $\rho$ is the local coarse-grained order parameter amplitude~\footnote{The XY spin degrees of freedom have fixed magnitude, $|S|=1$. We must define a local order
parameter whose amplitude can fluctuate which is achieved by coarse-graining.}. It is calculated as an average over the spins of the site at $\mathbf{x}_i$ and its six
nearest-neighbor sites:
\begin{equation}
  \rho(\mathbf{x}_i) = \frac17\left\lvert \epsilon_i\mathbf{S}_i + \sum_j^{n.n.} \epsilon_j\mathbf{S}_j\right\rvert.
\end{equation}

To derive a scaling form for the scalar susceptibility, we adapt the derivations given in Refs.~\cite{PodolskySachdev12,CrewseVojta21} to the classical case. Consider the
classical action for an order parameter $\psi$ in $d$ dimensions:
\begin{equation}
  S=\int\diff^d x \left[(\partial_{\mathbf{x}}\psi)^2 + (t + \delta t(\mathbf{x}))\psi^2 + u\psi^4\right]
  \label{eq:action}
\end{equation}
where $\delta t(\mathbf{x})$ represents a quenched random-mass disorder and $u$ is the quartic interaction
strength. The free energy density is defined as
\begin{equation}
  f = -\frac{1}{\beta V}\ln Z = -\frac{1}{\beta V}\ln\int D[\psi]e^{-S}
\end{equation}
where $V=L^d$ is the number of lattice sites. Taking the second derivative with respect to distance from criticality, we find
\begin{equation}
  \begin{split}
    \frac{\diff^2 f}{\diff t^2} &= \frac{1}{\beta V} \int \diff^d x \diff^d x' \\
                                &\times \left[\langle\psi^2(\mathbf{x})\psi^2(\mathbf{x'})\rangle - \langle\psi^2(\mathbf{x})\rangle\langle\psi^2(\mathbf{x'})\rangle\right]
  \end{split}
\end{equation}
which is precisely the $\mathbf{q}=0$ component of the susceptibility $\tilde{\chi}_{\rho^2\rho^2}$ of the square of the order parameter amplitude. Because the \emph{local} order parameter amplitude $\langle \rho (\mathbf{x}) \rangle$ is nonzero on both sides of the transition and at  criticality, the susceptibilities $\tilde{\chi}_{\rho^2\rho^2}$ and $\tilde{\chi}_{\rho\rho}$ are expected to have identical scaling behavior
\footnote{This can be shown explicitly by decomposing the local amplitude into average and fluctuations, $\rho(\mathbf{x}) = \langle\rho(\mathbf{x})\rangle + \delta \rho(\mathbf{x})$, and expanding both susceptibilities to leading order in $\delta \rho$.}.

The singular part of the free energy density fulfills the scaling form
\begin{equation}
  f(t) = b^{-d}f(tb^{1/\nu})
\end{equation}
for arbitrary length scale factor $b$. Now taking the second derivative of the free energy, we find
\begin{equation}
  f''(t) = b^{-d+2/\nu}f''(tb^{1/\nu})
\end{equation}
which implies the scaling form
\begin{equation}\label{eq:chi_scaling_form}
  \tilde{\chi}_{\rho\rho}(t,q) = b^{-d+2/\nu}\tilde{\chi}_{\rho\rho}(tb^{1/\nu},qb).
\end{equation}
If we set $b=q^{-1}$, the above equation transforms into
\begin{equation}\label{eq:chi_scaling_func}
  \tilde{\chi}_{\rho\rho}(t,q) = q^{d-2/\nu}X(tq^{-1/\nu})
\end{equation}
where $X(qt^{-\nu})$ is the scaling function of the scalar susceptibility. At criticality ($t=0$),
\begin{equation}\label{eq:chi_nu}
  \tilde{\chi}_{\rho\rho}(0,q) \sim q^{d-2/\nu}
\end{equation}
for small $q$. Correspondingly, the scalar correlation function at criticality is expected to be long-ranged, and $\chi_{\rho\rho}(0,\mathbf{x}) \sim
|\mathbf{x}|^{2/d-2/\nu}$, which implies that the scalar correlation function has the scaling form
\begin{equation}
  \chi_{\rho\rho}(t,\mathbf{x}) = |\mathbf{x}|^{2/\nu-2d} Y(t|\mathbf{x}|^{1/\nu})
  \label{eq:corrfunc_scaling}
\end{equation}
where $Y(t|\mathbf{x}|^{1/\nu})$ is the scaling function of the correlation function.

Since the correlation function is isotropic we will concern ourselves only with its dependence on the $x$ coordinate of the distance vector $\mathbf{x}=(x,y,z)$. We integrate (\ref{eq:corrfunc_scaling}) over $y$ and $z$ from $-\infty$ to $\infty$ (effectively setting the wave numbers $q_y=q_z=0$).
This increases the scale dimension of $\chi_{\rho\rho}$ by two:
\begin{equation}\label{eq:chi_form_eff} 
  \left.\vphantom{\sum}\chi_{\rho\rho}(t,x)\right\rvert_{q_y=q_z=0} = x^{2/\nu-2d+2} \bar Y(t x^{1/\nu}) ~.
\end{equation}
In the disordered phase above $T_c$, the scaling function $\bar Y$ is expected to decay as
$e^{-x /\xi_H}$ for large $x$, where $\xi_H$ is the amplitude correlation length. In contrast, the correlation function is expected to display an algebraic decay in the long-range ordered phase below $T_c$, described by $\chi_{\rho\rho}(t,x)|_{q_y=q_z=0} \sim x^{-4}$, which stems from the coupling of the amplitude mode to the massless Goldstone modes \cite{PodolskyAuerbachArovas11,PodolskySachdev12}.
Note, however, that a Monte Carlo study of the amplitude mode at the superfluid-insulator quantum phase transition \cite{GazitPodolskyAuerbach13} did not find conclusive evidence for the corresponding $\tau^{-4}$ decay of the scalar correlation function in imaginary time. Instead, their data followed an exponential form, as in the disordered phase,
suggesting that the asymptotic algebraic decay can only be observed at large distances inaccessible by the simulations.

We emphasize that eqs.~(\ref{eq:chi_scaling_form}) to (\ref{eq:chi_form_eff}) apply to the singular, critical part of $\chi_{\rho\rho}$. In addition, $\chi_{\rho\rho}$
is expected to have a non-critical background part which may need to be included in the analysis of the Monte Carlo data.

\section{Monte-Carlo Simulation}\label{sec:sim}

We studied our model using Monte Carlo simulations that combine conventional Metropolis single-spin updates~\cite{Metropolisetal53} with the Wolff cluster
algorithm~\cite{Wolff89}, which compliment each other well. The Wolff algorithm greatly reduces the critical slowing down, while the Metropolis algorithm equilibrates
the isolated sites or small clusters missed by the Wolff algorithm.

We simulated systems with a variety of different dilution concentrations, including $p=0$, $0.2$, $0.3$, $1/3$, and $0.5$, though we will be focusing on just $p=0$ and
$p=1/3$ in the following. For studying the Binder cumulant and to find $T_c$, we simulated several smaller systems (with sizes between $L=10$ and $L=100$) over a
relatively wide range of temperatures. For studying the correlation function and the scalar susceptibility, we simulated only systems with $L=128$ over a narrower range of temperatures near $T_c$.

\begin{figure}
  \includegraphics{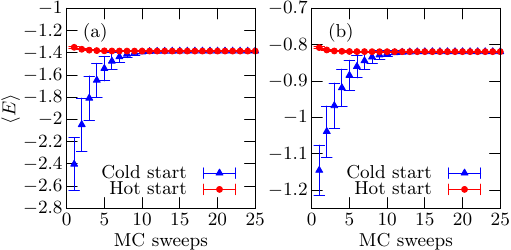}
  \caption{Comparison of the energy (averaged over 2000 disorder configurations) as a function of the Monte Carlo time (number of Monte Carlo sweeps) for hot and cold starts. All systems
  are of size $L=128$ and at distance from criticality $T-T_c=-0.18$ (below $T_c$). (a) Clean case, $p=0$, $T=2.022$. (b) Diluted case, $p=1/3$,
  $T=1.115$.}\label{fig:sweeps}
\end{figure}

Simulations started with 100 sweeps to equilibrate the system (one full Monte Carlo sweep being a Metropolis sweep followed by a Wolff sweep), then 500 sweeps to measure
the system, with one measurement per sweep. The number of equilibration sweeps was determined to be more than sufficient using the usual method of comparing hot and cold
simulation starts (see Fig.~\ref{fig:sweeps}).

Because of the random distribution of vacancies in each system, it is necessary to simulate many different disorder configurations (samples) in order to obtain a
statistically representative ensemble. Our data was averaged over 2000 to 8000 disorder configurations, depending on the system size and disorder strength.

\section{Results}\label{sec:results}
\subsection{Confirmation of thermodynamic critical behavior}\label{sec:binder}
\begin{figure}
  \includegraphics{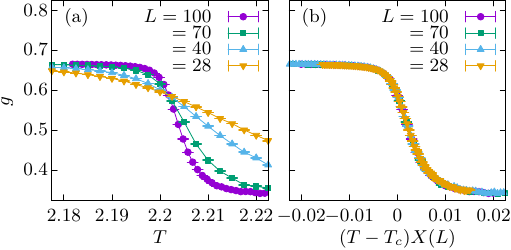}
  \caption{\label{fig:binder} (a) Binder cumulant $g$ as a function of temperature $T$ for $p$ = 0. (b) Scaling collapse of the Binder cumulant data.}
\end{figure}

To analyze the thermodynamics of our system, we first determine the critical temperature $T_c$ for each dilution strength $p$ by finding the crossing of the Binder
cumulant curves between several system sizes. In general, the crossings are of good quality and feature little drift between different system sizes, even at higher
dilutions. Though the smallest system sizes do tend to cross at noticeably lower $T$, increasing system size quickly decreases this drift to levels easily attributed to
Monte Carlo noise. Having found $T_c$, we then collapse the Binder cumulant by rescaling the distance from criticality $T-T_c$ for each curve by a constant factor $X(L)$
determined numerically. An example of the Binder crossing and rescaling for $p=0$ can be seen in Fig.~\ref{fig:binder}.

As previously discussed in Sec.~\ref{sec:model}, scaling predicts that the computed scale factors follow the power-law (\ref{eq:powerlaw_form}), so the value of $\nu$ is
then calculated by making numerical fits to this form. In addition, to fully confirm the system's thermodynamic critical behavior, values for the critical exponent
$\beta/\nu$ are calculated by analyzing the system-size dependence of the magnetization at criticality and making fits to the form
$\left.\langle|\mathbf{m}|\rangle\right\rvert_{T=T_c}\sim L^{-\beta/\nu}$.

\begin{figure}
  \includegraphics{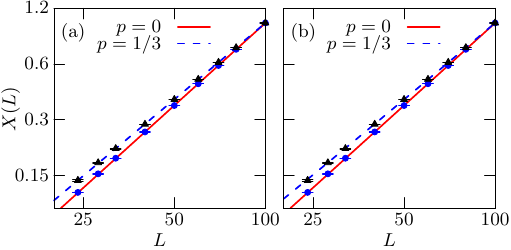}
  \caption{\label{fig:x_vs_l} Double logarithmic plots of the Binder collapse scale factors $X$ as a function of linear system size $L$.
    Error bars are comparable to the size of the symbols in each plot.
    (a) Fits to the pure power-law form (\ref{eq:powerlaw_form}) where $\nu$ is a fit parameter.
    (b) Fits of the same data to the functional form (\ref{eq:corrected_form}) that includes a correction-to-scaling term, with $\nu$ fixed at 0.6717.}
\end{figure}

Fig.~\ref{fig:x_vs_l}(a) shows fits of the scale factors $X(L)$ to (\ref{eq:powerlaw_form}) for $p=0$ and $p=1/3$. The clean data ($p=0$) follow a power law with an exponent very close to the
expected value $\nu=0.6717$. In contrast, the power law fit for $p=1/3$ gives an exponent above the expected value. This can be attributed to significant
deviations from the asymptotic behavior caused by finite-size effects, as discussed in Sec.~\ref{sec:model}. The values of $\nu$ and $\beta/\nu$ resulting from pure
power-law fits for each $p$, as well as the corresponding $T_c$, can be found in Table \ref{tab:table1}. These exponents should be understood as effective
(scale-dependent) exponents.

\begin{table}
  \caption{\label{tab:table1} Critical temperatures and effective (scale-dependent) values for $\nu$ and $\beta/\nu$ for each dilution concentration tested. Listed
  error bars account for statistical effects only.}
  \begin{ruledtabular}
    \begin{tabular}{llll}
      \multicolumn{1}{c}{$p$} & \multicolumn{1}{c}{$T_c$} & \multicolumn{1}{c}{$\nu$} & \multicolumn{1}{c}{$\beta/\nu$} \\
      \colrule
      0   & 2.20181(7) & 0.6760(7) & 0.5131(6) \\
      0.2 & 1.6713(6)  & 0.716(1) & 0.484(3) \\
      0.3 & 1.3911(5)  & 0.728(1) & 0.490(4) \\
      1/3 & 1.2947(4)  & 0.729(1) & 0.474(4) \\
      0.5 & 0.7822(5)  & 0.740(2) & 0.472(4) \\
    \end{tabular}
  \end{ruledtabular}
\end{table}

To account for the finite-size effects, the functional form (\ref{eq:powerlaw_form}) is modified with a multiplicative correction-to-scaling term:
\begin{equation}\label{eq:corrected_form}
  X(L) = aL^{1/\nu}(1+bL^{-\omega})
\end{equation}
where $a$ and $b$ are fit parameters and $\nu$ is fixed at $0.6717$. Preliminary fits for individual dilution values $p$ using the irrelevant exponent $\omega$ as a fit parameter
gave values of $\omega\approx 0.4$. As $\omega$ is expected to be universal, i.e., independent of the dilution, we fix $\omega$ at=0.4 for all $p$ in the final data analysis.
Fits to this corrected functional form for $p=0$ and $p=1/3$ can be
found in Fig.~\ref{fig:x_vs_l}(b). The fits are of reasonable quality, giving reduced $\chi^2$-values of $\chi^2/n=1.11$ for $p=0$ and $\chi^2/n=1.24$ for $p=1/3$.
Over our limited system-size range, the pure power-law fits and the fits including corrections to scaling in Fig.~\ref{fig:x_vs_l} are difficult to distinguish. This is caused by the small value of the irrelevant exponent $\omega$ that leads to a comparatively weak curvature in the curves produced by the functional form (\ref{eq:corrected_form}). The fits demonstrate, however, that our data for the scale factors $X$  for both $p=0$ and 1/3 are fully compatible with the clean critical behavior.

Analogous results were obtained for the exponent $\beta/\nu$ by making fits to the corrected form $\left.\langle|\mathbf{m}|\rangle\right\rvert_{T=T_c}=aL^{-\beta/\nu}(1+bL^{-\omega})$, with
$\beta/\nu$ fixed at the expected value of $0.519$~\cite{CampostriniHasenbuschPelissettoVicari06} and $\omega$ fixed at $0.4$, as before. From this we can conclude that
the diluted XY model features the same thermodynamic critical behavior as the clean case, agreeing with the prediction of the Harris criterion and previous results from the work of \citet{Santos-FilhoPlascak11}.

\subsection{Amplitude correlation function and correlation lengths}
\begin{figure}
  \includegraphics{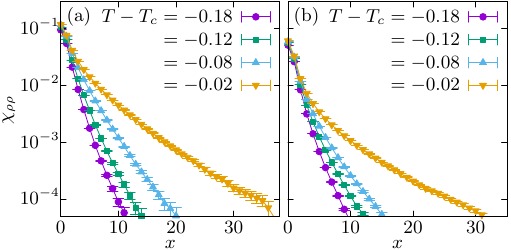}
  \caption{\label{fig:chi_vs_x} Semi-logarithmic plots of the correlation function $\chi_{\rho\rho}$ below $T_c$ as a function of separation $x$ (measured in units of
    the lattice constant) at $q_y=q_z=0$. (a) Clean case ($p = 0$). (b) Diluted case ($p = 1/3$).}
\end{figure}

We now turn to the main topic of this paper, our analysis of the scalar correlation function and the scalar susceptibility, which characterize the amplitude fluctuations. Figure \ref{fig:chi_vs_x} shows the scalar correlation function for several temperatures below $T_c$.
The curves feature the expected behavior: correlations drop off quickly with increasing $x$, while the correlation lengths increases as the system approaches $T_c$. The diluted case behaves analogously to the clean case, only with slightly shorter range correlations, as expected.

\begin{figure}
  \includegraphics{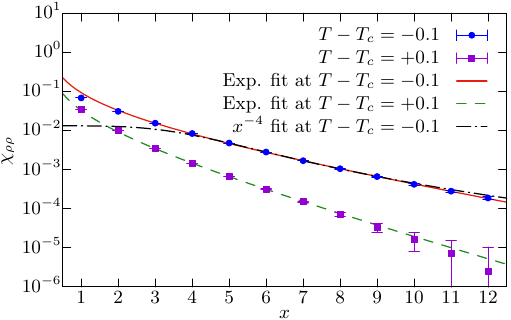}
  \caption{\label{fig:chi_fit_ex} Fits of the scalar correlation function to functional form $\chi_{\rho\rho}(t,x)|_{q_y=q_z=0}  \sim x^{2/\nu-2d+2} e^{-x /\xi_H}$ (denoted ``Exp. fit'' in figure key) for $p=0$ and $T-T_c=-0.1$
  ($T=2.102$) and $T-T_c=+0.1$ ($T=2.302$), with the exponent $\nu$ fixed at the asymptotic value $\nu=0.6717$, as well as a fit to the form $a/(b+x^4)$ (denoted ``$x^{-4}$ fit'' in figure key) for $T-T_c=-0.1$. The fits exclude data at short distances ($x\lesssim5$) when the scaling form (\ref{eq:chi_form_eff}) is not expected to hold.}
\end{figure}

Let us now test whether the scalar correlation function $\chi_{\rho\rho}$ fulfills the predictions derived in Sec.~\ref{sec:scalar} by fitting the numerical data for $\chi_{\rho\rho}$ at $q_y=q_z=0$ to the scaling form (\ref{eq:chi_form_eff}). For temperatures above $T_c$, the scaling function is expected to decay exponentially for large $x$. We therefore fit the data to $\chi_{\rho\rho}(t,x)|_{q_y=q_z=0}  \sim x^{2/\nu-2d+2} e^{-x /\xi_H}$. An example of such a fit is presented in Fig.\ \ref{fig:chi_fit_ex} for $p=0$ and $T-T_c =0.1$.
The figure shows that the data are in good agreement with the prediction. In the long-range ordered phase at $T < T_c$, the scalar correlation function is expected to fall off, in the large-$x$ limit, like $x^{-4}$ instead of exponentially, as discussed in Sec.~\ref{sec:scalar}. However, a fit of the data for $T-T_c=-0.1$ in Fig.\ \ref{fig:chi_fit_ex} to the same exponential form as in the disordered phase above $T_c$ is of somewhat better quality than the fit to the power-law tail. This resembles the results for the superfluid-insulator quantum phase transition \cite{GazitPodolskyAuerbach13} where the data for the (imaginary time) scalar correlation function in the ordered phase followed an exponential decay rather than the expected $\tau^{-4}$ power law. The discrepancy suggests that the true asymptotic behavior can only be observed at larger distances that are unreachable in our simulations due to size limitations and Monte Carlo noise.

\begin{figure}
  \includegraphics{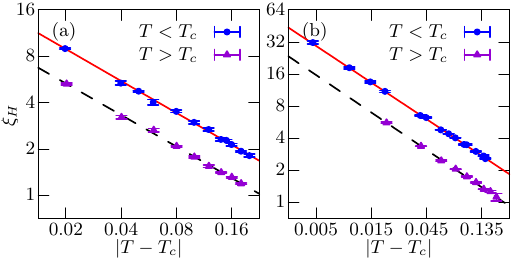}
  \caption{Double logarithmic plots of the correlation lengths $\xi_H$ (for systems both above and below $T_c$) fitted to the functional form (\ref{eq:xi_t}). The width of the error bars are comparable to the
    size of the points. (a) Clean case ($p=0$); fits yielded $\nu=0.688$ below $T_c$ and $\nu=0.680$ above $T_c$. (b) Diluted case ($p=1/3$); fits yielded $\nu=0.721$ below $T_c$ and $\nu=0.736$ above $T_c$.}\label{fig:xi_vs_t}
\end{figure}

The dependence of the amplitude correlation length $\xi_H$ resulting from the above fits on the distance from criticality $t$ is presented in Fig.~\ref{fig:xi_vs_t}.
The data for both the clean case and the diluted case follow the power law
\begin{equation}\label{eq:xi_t}
  \xi_H\sim |t|^{-\nu}
\end{equation}
in good approximation. In the
clean case, the value of $\nu$ resulting from a fit to eq.\ (\ref{eq:xi_t}) is close to the expected value of $\nu=0.6717$. As in Sec.~\ref{sec:binder}, the value for non-zero dilution is
larger. As alluded to in Sec.~\ref{sec:model}, these deviations from the expected value of $\nu$ are the result of finite-size effects, which are very pronounced because
the Harris criterion is only barely fulfilled. It is also worth noting that the effective values for $\nu$ resulting from $\xi_H$ are similar to those found in the
analysis of the Binder cumulant in Sec.~\ref{sec:binder}.

\subsection{Scaling behavior of the scalar susceptibility}
\begin{figure}
  \includegraphics{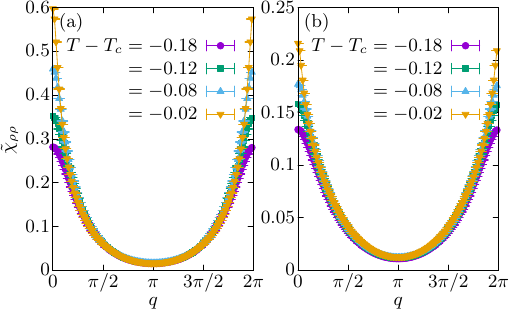}
  \caption{Scalar susceptibility $\tilde{\chi}_{\rho\rho}$ below $T_c$ as a function of wave number $q$.
  (a) Clean case ($p=0$). (b) Diluted case ($p=1/3$).}\label{fig:chi_vs_q}
\end{figure}

\begin{figure}
  \includegraphics{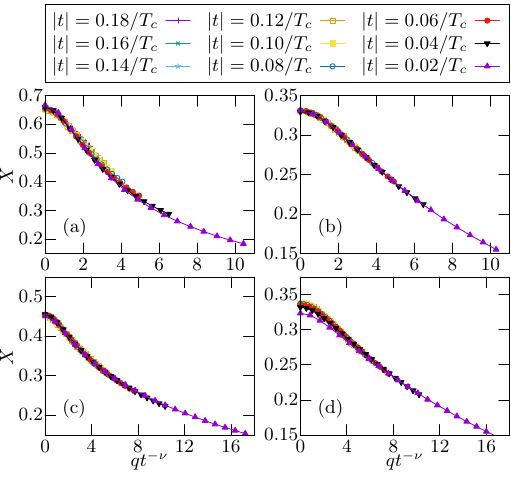}
  \caption{Scaling collapse of $\tilde \chi_{\rho\rho}$ above and below the critical temperature. Shown is $X$, defined in (\ref{eq:X}), vs. the scaling coordinate $qt^{-\nu}$. The width of the
    error bars are comparable to the size of the points.
    Collapses for $p=0$ used $\nu=0.676$ and those for $p=1/3$ used $\nu=0.729$.
    (a) $p=0$, below $T_c$.
    (b) $p=0$, above $T_c$.
    (c) $p=1/3$, below $T_c$.
    (d) $p=1/3$, above $T_c$.}\label{fig:X_all}
\end{figure}

Figure \ref{fig:chi_vs_q} shows the scalar susceptibility $\tilde{\chi}_{\rho\rho}$ in the long-range ordered phase for both the clean case ($p=0$) and the diluted case ($p=1/3$).
We now seek a scaling collapse of  $\tilde{\chi}_{\rho\rho}(t,q)$ in order to fully determine whether or not the amplitude fluctuations violate naive scaling. The scaling form
of $\tilde{\chi}_{\rho\rho}$  was given in (\ref{eq:chi_scaling_form}). To fit the Monte Carlo data, we need to include a non-critical background term which we approximate as a constant $\tilde{\chi}_0$ and treat it
as a fit parameter. This leads to the form $\tilde{\chi}_{\rho\rho}(t,q) = t^{d\nu-2} X(qt^{-\nu}) +\tilde{\chi}_0$. To test whether the Monte Carlo results fulfill this
scaling form, we attempt to collapse
\begin{equation}\label{eq:X}
  X(qt^{-\nu}) = t^{2-d\nu} [\tilde{\chi}_{\rho\rho}(t,q) - \tilde{\chi}_0]
\end{equation}
for different distances from criticality onto a common master curve. Figure~\ref{fig:X_all} shows that reasonably good collapses can be achieved for both $p=0$ and
$p=1/3$, using the same effective $\nu$ values found in the analysis of the Binder cumulant.
If the asymptotic value of $\nu=0.6717$ is used, the quality of the fits is lower.
Once again, these deviations from the expected value of $\nu=0.6717$ are the results of finite size effects, and, in the diluted system, of the slow
renormalization of the disorder strength. Small deviations from perfect data collapse can also be attributed to our simple approximation of the non-critical part of
$\chi_{\rho\rho}$ and to the uncertainty of $T_c$ \footnote{In Fig.~\ref{fig:X_all}(d) it is clear that the curve closest to criticality, $|t|/T_c=0.02$, is significantly offset
  from the other curves. One possible explanation for this inaccuracy is that our value for the critical temperature for $p=1/3$ could be lacking precision. This would
  most strongly affect the scaling of the data close to $T_c$.}.

\section{Conclusion}\label{sec:conc}
To summarize, we have studied the order-parameter amplitude fluctuations in a site-diluted three-dimensional classical XY model. This was motivated by the unconventional
localization behavior of the amplitude (Higgs) mode recently observed near the superfluid-Mott glass quantum phase transition of disordered bosons, modeled by a
(2+1)-dimensional quantum rotor model \cite{PuschCrewseHoyosVojta20,CrewseVojta21,PuschGetelinaHoyosVojta21}. In the absence of disorder, the transitions in the classical XY model and the quantum rotor model
are described by the same field theory and belong to the same universality class. However, the disorder is perfectly correlated in the imaginary time direction in the
quantum case whereas it is uncorrelated in all directions in the classical case. As a result, disorder turns out to be an irrelevant perturbation in the classical case
while it changes the critical behavior in the quantum case \cite{VojtaCrewsePuschmannArovasKiselev16}. Comparing the properties of the amplitude fluctuations in the two cases can thus help us to
disentangle possible reasons for the amplitude mode localization at the superfluid-Mott glass transition \footnote{The amplitude mode localization at the
superfluid-Mott glass transition was mainly discussed in terms of the real-frequency spectral densities of the scalar susceptibility. However, Ref.~\cite{CrewseVojta21}
demonstrated that the unconventional behavior is already clearly visible in the imaginary time Monte Carlo data that can be directly compared to the classical Monte
Carlo results in the present paper.}.

The present Monte Carlo results for the scalar susceptibility of the site-diluted classical XY model do not show any traces of unconventional behavior. In contrast to
the quantum case, they agree with predictions of (naive) scaling theory. We have confirmed this by analyzing the functional form of the scalar (amplitude-amplitude)
correlation function in real space as well as by achieving a scaling collapse of the scalar susceptibility in momentum space. While the data presented in the previous
sections focused on the dilution value $p=1/3$, we have obtained analogous results for the other studied values including the strongest dilution of $p=1/2$.

We now return to the question raised in the introduction and discuss what these results can tell us about the reasons for the  unconventional behavior of the scalar susceptibility
at the superfluid-Mott glass quantum phase transition \cite{PuschCrewseHoyosVojta20,CrewseVojta21,PuschGetelinaHoyosVojta21}. If the amplitude model localization simply was a manifestation
of Anderson localization due to the randomness in the bare Hamiltonian or, equivalently, in the bare field theory, one might expect similar localization behavior in the classical XY model studied in
the present paper. This stems from the fact that the eigenmodes of the (Gaussian part of the) bare classical field theory (\ref{eq:action}) are expected to be localized due to the
spatially uncorrelated random mass disorder. (Even in three dimensions and for weaker disorder, this is expected to hold for the eigenstates having the lowest eigenvalues that dominate near the critical point.)
The results of the present paper suggest that this expectation may be too naive, and renormalization phenomena play a key role.

Importantly, the renormalizations of the amplitude fluctuations
in the classical and quantum cases differ significantly from each other. To understand this, let us compare the scaling forms of the scalar
susceptibility in the classical and quantum cases. According to eq.~(\ref{eq:chi_scaling_form}), the scale dimension of the scalar susceptibility of the classical XY model is $-d+2/\nu$.
Using $d=3$ and $\nu=0.6717$ (which applies to both the undiluted and the diluted classical XY models) gives a scale dimension very close to zero. This implies that the
critical (scaling) part of the scalar susceptibility remains essentially unchanged under renormalization.
In contrast, the scale dimension of the scalar susceptibility at the superfluid-Mott glass transition is given by $-(d+z)+2/\nu$ (see eq.~(12) of Ref.~\cite{CrewseVojta21}).
Using $d=2$ together with the critical exponent values $z=1.52$ and $\nu=1.16$ \cite{VojtaCrewsePuschmannArovasKiselev16} gives a negative scale dimension of about $-1.8$. Consequently, the
critical (scaling) part of the scalar susceptibility is expected to decrease rapidly under renormalization and to become negligible at low energies close to the
transition. The scalar response measured in Monte Carlo simulations or potential experiments would then stem from the noncritical background part of the scalar
susceptibility and thus be dominated by short-wavelength and/or localized (microscopic) degrees of freedom.

To further test this scenario one could study disordered quantum phase transitions in a different universality class, with exponent values that do not lead to a negative scale
dimension of the scalar susceptibility. At these transitions, the observed scalar response should be conventional and fulfill (naive) scaling. This has recently been confirmed
for a quantum phase transition of bosons on a random Voronoi-Delaunay lattice \cite{VishnuNarayananPuschmannVojta}. From a broader perspective, such investigations would help address the
question under what conditions disordered quantum phase transitions can exhibit exotic real-time dynamics even if their thermodynamics is conventional. Is it possible to
classify dynamical phenomena in a similar manner as the thermodynamic (quantum) critical behavior \cite{Vojta06,VojtaHoyos14}?

\begin{acknowledgments}
This work was supported in part by the National Science Foundation under Grants No.\ DMR-1828489 and No.\ OAC-1919789.
\end{acknowledgments}

\bibliographystyle{apsrev4-2}
\bibliography{refs.bib}

\end{document}